\documentclass{article}
\usepackage{spconf,amsmath,graphicx,hyperref}

\usepackage{amssymb}
\usepackage{amsmath}
\usepackage{float}
\usepackage{stfloats}
\usepackage[table]{xcolor}
\usepackage{graphicx}
\usepackage{booktabs}
\usepackage{url}
\usepackage{multirow}
\usepackage{caption}
\usepackage{subcaption}
\usepackage{makecell}
\usepackage{setspace}

\title{XM-ALIGN: Unified Cross-Modal Embedding Alignment for Face-Voice Association}

\name{Zhihua Fang$^{1,2}$ \qquad Shumei Tao$^{1,2}$ \qquad Junxu Wang$^{3}$ \qquad Liang He$^{1,2,4,5,\dagger}$
\thanks{
$^\dagger$Corresponding author (heliang@mail.tsinghua.edu.cn).
This work was supported by the National Natural Science Foundation of China under Grant No. 62366051.
}}

\address{
$^{1}$School of Computer Science and Technology, Xinjiang University, Urumqi 830017, China\\
$^{2}$Xinjiang Multimodal Information Technology Engineering Research Center, Urumqi 830017, China\\
$^{3}$Urumqi Branch, China Mobile Group Xinjiang Co., Ltd, Urumqi 830018, China\\
$^{4}$School of Intelligence Science and Technology, Xinjiang University, Urumqi 830017, China\\
$^{5}$Department of Electronic Engineering, Tsinghua University, Beijing 100084, China
}

\begin{document}
\ninept
\maketitle
\begin{abstract}
This paper introduces our solution, XM-ALIGN (Unified Cross-Modal Embedding Alignment Framework), proposed for the FAME challenge at ICASSP 2026. Our framework combines explicit and implicit alignment mechanisms, significantly improving cross-modal verification performance in both ``heard'' and ``unheard'' languages. By extracting feature embeddings from both face and voice encoders and jointly optimizing them using a shared classifier, we employ mean squared error (MSE) as the embedding alignment loss to ensure tight alignment between modalities. Additionally, data augmentation strategies are applied during model training to enhance generalization. Experimental results show that our approach demonstrates superior performance on the MAV-Celeb dataset. The code will be released at \textcolor[rgb]{0.874, 0.0, 0.486}{\url{https://github.com/PunkMale/XM-ALIGN}}.
\end{abstract}
\section{Introduction}
Face and voice are the most fundamental features for distinguishing human identities~\cite{FAME2024}. Although unimodal identity verification systems have achieved significant success, the challenge of performing cross-modal identity verification remains. In the FAME 2026 challenge~\cite{FAME2026}, domain shifts caused by language variations introduce additional complexity. The key challenge lies in ensuring that the learned identity embeddings maintain robustness across both ``heard'' and ``unheard'' languages. Traditional approaches~\cite{SSNet2020ICASSP, FOP2022ICASSP, SBNet2023ICASSP,PAEFF2025INTERSPEECH} typically rely on complex gated fusion mechanisms or separate projection spaces, which may inadvertently retain modality- or language-specific noise, thus hindering the model’s generalization ability in unseen-unheard scenarios.

To address this challenge, we propose a unified cross-modal embedding alignment (\textbf{XM-ALIGN}) framework, which aims to improve the performance of cross-modal face-voice association systems in multilingual environments by combining explicit and implicit alignment mechanisms. Our approach consists of two core components: a face encoder and a voice encoder, which extract feature embeddings from facial images and speech signals, respectively. By introducing an embedding alignment loss, we explicitly ensure that the face and voice embedding spaces are more consistent, thereby effectively reducing alignment errors caused by modality differences. Additionally, through the design of a shared classification head, we implicitly strengthen the mutual alignment of face and voice features during optimization. Our method performs exceptionally well in the FAME 2026 challenge, particularly in cross-modal verification tasks with ``heard'' and ``unheard'' languages, demonstrating significant improvements.

\section{Methods}

\begin{figure}[tbp]
    \centering
    \includegraphics[width=1.0\linewidth]{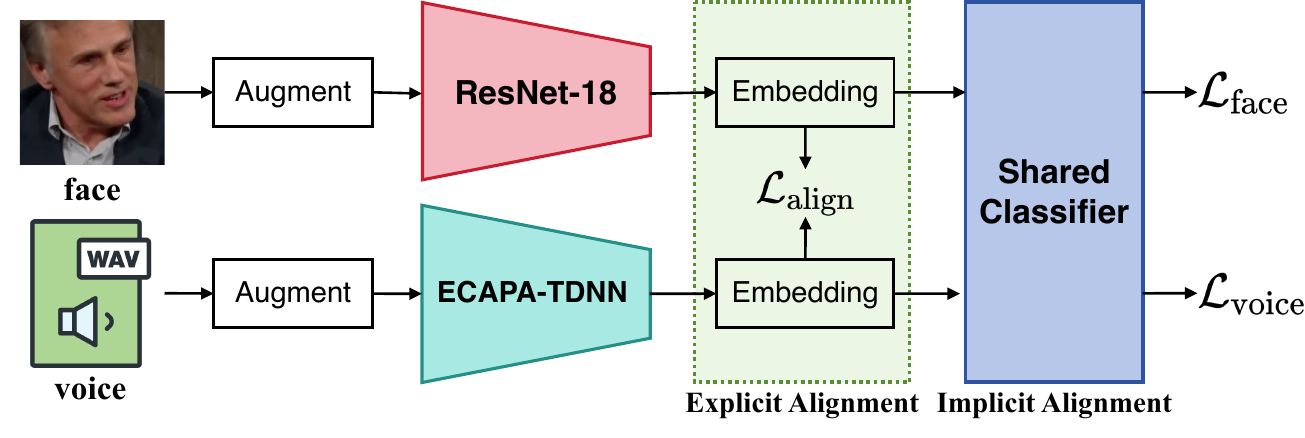}
    \caption{Our Proposed XM-ALIGN Framework.}
    \label{fig:xm-align-net}
\end{figure}

The overall architecture of XM-ALIGN is shown in Fig.~\ref{fig:xm-align-net}. The system consists of two parallel streams: a visual encoder \( E_\text{face}(\cdot) \) using a ResNet-18 backbone, and an audio encoder \( E_\text{voice}(\cdot) \) based on the ECAPA-TDNN architecture. Given a face image \( x_f \) and a voice segment \( x_v \) from the same identity, we extract their \( D \)-dimensional embeddings: \( \mathbf{e}_f = E_\text{face}(x_f) \) and \( \mathbf{e}_v = E_\text{voice}(x_v) \). The key idea is how to align the embeddings from different modalities during training.

\subsection{Explicit Alignment Based on Embedding Alignment}

To effectively align the feature embeddings of face and voice, we introduce an explicit alignment objective that forces the audio embedding to approach the visual embedding of the same identity in the feature space. We propose an embedding alignment loss (MSE loss), which explicitly adjusts the distance between the two embeddings by computing the mean squared error between the face and voice embeddings. The loss function is defined as follows:
\begin{equation}
    \mathcal{L}_{\text{align}} = \frac{1}{N} \sum_{i=1}^{N} \| \mathbf{e}_{f}^{(i)} - \mathbf{e}_{v}^{(i)} \|_2^2,
\end{equation}
where \( \mathbf{e}_{f}^{(i)} \) and \( \mathbf{e}_{v}^{(i)} \) represent the face and voice embeddings of the \( i \)-th sample, and \( N \) is the batch size. By minimizing this loss function, we ensure that the distance between the face and voice embeddings is minimized for the same identity, thereby effectively enhancing the cross-modal association performance.

\subsection{Implicit Alignment Based on Shared Classifier}
In addition to the explicit alignment of embeddings, we also design a shared classifier~\cite{ShareFC2021ICASSP} to implicitly align the face and voice embeddings. In our design, the classifier simultaneously receives embeddings from both the face and voice encoders and performs joint optimization in a shared space. Specifically, we employ a single weight matrix $\mathbf{W} \in \mathbb{R}^{C \times D}$ (where $C$ is the number of identities) to classify both $\mathbf{e}_f$ and $\mathbf{e}_v$ as follows:
\begin{equation}
    \left\{
    \begin{array}{l}
    \mathcal{L}_{\text{face}} = -\frac{1}{N} \sum_{i=1}^{N} \log \frac{e^{\mathbf{W}_{y_i}^T \mathbf{e}_f^{(i)}}}{\sum_{j=1}^{C} e^{\mathbf{W}_{j}^T \mathbf{e}_f^{(i)}}}, \\
    \mathcal{L}_{\text{voice}} = -\frac{1}{N} \sum_{i=1}^{N} \log \frac{e^{\mathbf{W}_{y_i}^T \mathbf{e}_v^{(i)}}}{\sum_{j=1}^{C} e^{\mathbf{W}_{j}^T \mathbf{e}_v^{(i)}}},
    \end{array}
    \right.
  \end{equation}
where $y_i$ is the true identity label, and $\mathbf{W}_{j}$ represents the $j$-th column of the shared weight matrix. This parameter sharing acts as a strong regularizer. By applying the same decision boundary to both modalities, the encoders are implicitly constrained to map face and voice inputs to the same semantic region in the feature space.

\subsection{Overall Training Objective}
The final training objective of XM-ALIGN is the weighted sum of identity discrimination loss and modality alignment loss. By jointly optimizing these terms, the embeddings learned by the model are both discriminative (separated by identity) and cohesive (insensitive to modality). The total loss function is expressed as:
\begin{equation}
    \mathcal{L}_{\text{total}} = \mathcal{L}_{\text{face}} + \mathcal{L}_{\text{voice}} + \lambda \mathcal{L}_{\text{align}},
\end{equation}
where \( \mathcal{L}_{\text{face}} \) and \( \mathcal{L}_{\text{voice}} \) use cross-entropy (CE) loss, and \( \lambda \) is a hyperparameter that balances the strength of the explicit alignment constraint. During the inference phase, the similarity between face and voice is directly calculated as the cosine similarity between their embeddings, without the need for an additional fusion network.

\section{Experimental Setup and Results}

\subsection{Experimental Setup}
We conducted extensive experimental evaluations on the MAV-Celeb~\cite{mavceleb} dataset to assess the performance of the XM-ALIGN framework. We used ECAPA-TDNN~\cite{ecapa_tdnn2020interspeech} with 1024 channels as the voice encoder, ResNet-18~\cite{deep_residual2016cvpr} as the face encoder, and set the embedding dimension to 512. Cross-entropy was used as the classification loss, and mean squared error (MSE) was used as the embedding alignment loss. To enhance the model's generalization capability, data augmentation was applied: speech augmentation included reverberation~\cite{rirs2017icassp} and noise injection~\cite{musan2015}, while image augmentation included random horizontal flipping, Gaussian blurring, and grayscale conversion. The model was trained for a total of 500 epochs, with an initial learning rate of 0.001 and a decay rate of 0.97 per epoch, optimized using the Adam optimizer. The final model performance was determined by the average weights of the last 5 epochs. The evaluation metric used was equal error rate (EER).

\begin{table}[tbp]
    \centering
    \caption{EER (\%) results on MAV-Celeb using different classifier configurations and embedding alignment strategies. The best two results are marked in bold.}
    \label{tab:results}
    \resizebox{\linewidth}{!}{%
    \begin{tabular}{c c c c c c c c c c }
    \toprule
    \multicolumn{10}{c}{\textbf{MAV-Celeb V1}} \\
    \midrule
    \textbf{\#} & 
    \makecell{\textbf{Cls.} \\ \textbf{Head}} & 
    \makecell{\textbf{Cls.} \\ \textbf{Loss}} & 
    \makecell{\textbf{Align.} \\ \textbf{Metric}} & 
    \makecell{\textbf{Align.} \\ \textbf{Weight} \\ ($\boldsymbol{\lambda}$)} & 
    \makecell{\textbf{English} \\ \textbf{heard}} & 
    \makecell{\textbf{English} \\ \textbf{unheard}} & 
    \makecell{\textbf{Urdu} \\ \textbf{heard}} & 
    \makecell{\textbf{Urdu} \\ \textbf{unheard}} & 
    \makecell{\textbf{Overall} \\ \textbf{Score}} \\
    \midrule
    \textbf{1} & \multicolumn{4}{c}{Baseline} & \textbf{29.3} & 40.4 & 25.8 & 37.9 & 33.4 \\
    \midrule
    \textbf{2} & Separate & AAM & MSE & 0.1 & 42.1 & 40.8 & 41.5 & 43.3 & 41.9 \\
    \textbf{3} & Separate & CE & Cosine & 1.0 & 38.2 & \textbf{36.7} & 25.0 & 29.5 & 32.3 \\
    \textbf{4} & Separate & CE & MSE & 0.1 & 35.3 & 40.0 & 24.6 & 25.0 & 31.2 \\
    \textbf{5} & Separate & CE & MSE & 1.0 & 35.2 & 39.1 & \textbf{24.1} & 25.0 & \textbf{30.8} \\
    \textbf{6} & Shared & CE & - & - & 43.0 & 40.0 & 37.5 & 37.9 & 39.6 \\
    \textbf{7} & Shared & CE & MSE & 0.1 & 35.7 & 40.4 & \textbf{22.8} & \textbf{24.1} & \textbf{30.8} \\
    \textbf{8} & Shared & CE & MSE & 1.0 & \textbf{34.2} & \textbf{38.2} & 25.4 & \textbf{24.6} & \textbf{30.6} \\
    \midrule
    \multicolumn{10}{c}{\textbf{MAV-Celeb V3}} \\
    \midrule
    \textbf{\#} & 
    \makecell{\textbf{Cls.} \\ \textbf{Head}} & 
    \makecell{\textbf{Cls.} \\ \textbf{Loss}} & 
    \makecell{\textbf{Align.} \\ \textbf{Metric}} & 
    \makecell{\textbf{Align.} \\ \textbf{Weight} \\ ($\boldsymbol{\lambda}$)} & 
    \makecell{\textbf{English} \\ \textbf{heard}} & 
    \makecell{\textbf{English} \\ \textbf{unheard}} & 
    \makecell{\textbf{German} \\ \textbf{heard}} & 
    \makecell{\textbf{German} \\ \textbf{unheard}} & 
    \makecell{\textbf{Overall} \\ \textbf{Score}} \\
    \midrule
    \textbf{9} & \multicolumn{4}{c}{Baseline} & 34.5 & 43.2 & 39.6 & 43.7 & 40.2 \\
    \midrule
    \textbf{10} & Separate & CE & MSE & 0.1 & 32.9 & 42.2 & 34.4 & 32.1 & 35.4 \\
    \textbf{11} & Separate & CE & MSE & 0.3 & 32.4 & 40.5 & 33.9 & \textbf{30.4} & 34.3 \\
    \textbf{12} & Separate & CE & MSE & 1.0 & 31.8 & 53.8 & 54.5 & \textbf{30.8} & 42.7 \\
    \textbf{13} & Shared & CE & MSE & 0.1 & \textbf{31.2} & 41.0 & \textbf{32.6} & 32.1 & \textbf{34.2} \\
    \textbf{14} & Shared & CE & MSE & 1.0 & 33.5 & \textbf{39.3} & 33.9 & 31.7 & 34.6 \\
    \midrule
    \textbf{15} & \multicolumn{4}{c}{Score Fusion (\#11,13,14)} & \textbf{32.4} & \textbf{38.7} & \textbf{32.1} & \textbf{30.8} & \textbf{33.5} \\
    \bottomrule
    \end{tabular}
    }
\end{table}

\subsection{Results}
As shown in Table~\ref{tab:results}, we conducted ablation experiments on the MAV-Celeb V1 dataset to validate the effectiveness of different components. On MAV-Celeb V1, we evaluated the impact of the classification loss function, showing that CE loss outperforms AAM-Softmax~\cite{arcface2019cvpr}, likely due to overfitting from the AAM margin penalty on small datasets. For embedding alignment, MSE outperforms cosine distance, highlighting the advantage of enforcing absolute proximity in Euclidean space. The significant gap between experiments \#6 and \#7 emphasizes the importance of explicit embedding alignment, which is necessary to bridge the modality gap. The best performance was achieved by combining a shared classification head (experiment \#8), demonstrating the value of combining implicit semantic and explicit geometric alignment.

These findings generalize to the V3 dataset. The shared classification head with MSE alignment (experiments \#13 and \#14) demonstrated strong performance, especially in unseen language challenges. While the best single model (experiment \#13) achieved an EER of 34.2\%, score fusion of the top three models (experiments \#11, \#13, and \#14) reduced the EER to 33.5\%, further demonstrating the effectiveness of multi-model ensembles in multilingual face-voice association.

\section{Conclusion}
This paper proposed the XM-ALIGN framework, which effectively improves performance in the multilingual face-voice association task by combining explicit and implicit alignment. In the FAME challenge at ICASSP 2026, experimental results show that using a shared classification head along with the embedding alignment strategy significantly enhances the model's accuracy in cross-lingual verification. Furthermore, through a simple score fusion strategy, we further improve the overall system performance, demonstrating the effectiveness of multi-model ensemble in multilingual environments.

\bibliographystyle{IEEEbib}
\bibliography{references}

\begin{thebibliography}{10}

\bibitem{FAME2024}
Muhammad~Saad Saeed, Shah Nawaz, Marta Moscati, Rohan~Kumar Das, Muhammad~Salman Tahir, Muhammad~Zaigham Zaheer, Muhammad~Irzam Liaqat, Muhammad~Haris Khan, Karthik Nandakumar, Muhammad~Haroon Yousaf, and Markus Schedl,
\newblock ``A synopsis of fame 2024 challenge: Associating faces with voices in multilingual environments,''
\newblock in {\em ACM MM}, 2024, pp. 11333--11334.

\bibitem{FAME2026}
Marta Moscati, Ahmed Abdullah, Muhammad~Saad Saeed, Shah Nawaz, Rohan~Kumar Das, Muhammad~Zaigham Zaheer, Junaid Mir, Muhammad~Haroon Yousaf, Khalid Malik, and Markus Schedl,
\newblock ``Face-voice association in multilingual environments (fame) 2026 challenge evaluation plan,'' 2025.

\bibitem{SSNet2020ICASSP}
Shah Nawaz, Muhammad~Kamran Janjua, Ignazio Gallo, Arif Mahmood, and Alessandro Calefati,
\newblock ``Deep latent space learning for cross-modal mapping of audio and visual signals,''
\newblock in {\em DICTA}, 2019, pp. 1--7.

\bibitem{FOP2022ICASSP}
Muhammad~Saad Saeed, Muhammad~Haris Khan, Shah Nawaz, Muhammad~Haroon Yousaf, and Alessio Del~Bue,
\newblock ``Fusion and orthogonal projection for improved face-voice association,''
\newblock in {\em ICASSP}, 2022, pp. 7057--7061.

\bibitem{SBNet2023ICASSP}
Muhammad~Saad Saeed, Shah Nawaz, Muhammad~Haris Khan, Muhammad Zaigham~Zaheer, Karthik Nandakumar, Muhammad~Haroon Yousaf, and Arif Mahmood,
\newblock ``Single-branch network for multimodal training,''
\newblock in {\em ICASSP}, 2023, pp. 1--5.

\bibitem{PAEFF2025INTERSPEECH}
Abdul Hannan, Muhammad~Arslan Manzoor, Shah Nawaz, Muhammad~Irzam Liaqat, Markus Schedl, and Mubashir Noman,
\newblock ``{PAEFF: Precise Alignment and Enhanced Gated Feature Fusion for Face-Voice Association},''
\newblock in {\em {INTERSPEECH}}, 2025, pp. 2710--2714.

\bibitem{ShareFC2021ICASSP}
Leda Sarı, Kritika Singh, Jiatong Zhou, Lorenzo Torresani, Nayan Singhal, and Yatharth Saraf,
\newblock ``A multi-view approach to audio-visual speaker verification,''
\newblock in {\em ICASSP}, 2021, pp. 6194--6198.

\bibitem{mavceleb}
Shah Nawaz, Muhammad~Saad Saeed, Pietro Morerio, Arif Mahmood, Ignazio Gallo, Muhammad~Haroon Yousaf, and Alessio Del~Bue,
\newblock ``Cross-modal speaker verification and recognition: A multilingual perspective,''
\newblock in {\em CVPRW}, 2021, pp. 1682--1691.

\bibitem{ecapa_tdnn2020interspeech}
Brecht Desplanques, Jenthe Thienpondt, and Kris Demuynck,
\newblock ``{ECAPA-TDNN: Emphasized Channel Attention, Propagation and Aggregation in TDNN Based Speaker Verification},''
\newblock in {\em INTERSPEECH}, 2020, pp. 3830--3834.

\bibitem{deep_residual2016cvpr}
Kaiming He, Xiangyu Zhang, Shaoqing Ren, and Jian Sun,
\newblock ``Deep residual learning for image recognition,''
\newblock in {\em CVPR}, 2016.

\bibitem{rirs2017icassp}
Tom Ko, Vijayaditya Peddinti, Daniel Povey, Michael~L. Seltzer, and Sanjeev Khudanpur,
\newblock ``A study on data augmentation of reverberant speech for robust speech recognition,''
\newblock in {\em ICASSP}, 2017, pp. 5220--5224.

\bibitem{musan2015}
David Snyder, Guoguo Chen, and Daniel Povey,
\newblock ``Musan: A music, speech, and noise corpus,'' 2015.

\bibitem{arcface2019cvpr}
Jiankang Deng, Jia Guo, Niannan Xue, and Stefanos Zafeiriou,
\newblock ``Arcface: Additive angular margin loss for deep face recognition,''
\newblock in {\em CVPR}, 2019, pp. 4685--4694.

\end{thebibliography}

\end{document}